# Albizu University & AGM University

School of Professional Studies and Cyberpsychology

# Cognitive AI framework 2.0: advances in the simulation of human thought

**Dr. Rommel Salas-Guerra, PhD, D.B.A.**

Professor of Cyberpsychology, Albizu University, Florida

Professor, Professional Studies School, AGM University, USA

January 2026

**Abstract**

The Human Cognitive Simulation Framework constitutes a paradigmatic advance in the integration of human cognitive capabilities into artificial intelligence systems, aiming to optimize personalization, adaptability, and long-term coherence in human–AI interaction. The framework integrates short-term memory (conversation context), long-term memory (interaction context), advanced cognitive processing, and governed knowledge management into a unified architectural model that ensures contextual continuity across sessions and controlled persistence of relevant information over time. A central contribution is the introduction of a unified memory architecture that synchronizes these contexts under explicit governance mechanisms, improving accessibility, traceability, and operational efficiency. The framework further incorporates differentiated cognitive processing modules—logical, creative, and analogical—conceptually inspired by the functional specialization of the cerebral hemispheres, enabling both structured task execution and complex contextual inference. Through dynamic and selective knowledge updating, the system enriches the pre-trained capabilities of large language models while preserving control over long-term information accumulation, thus supporting adaptive and personalized interaction in domains such as education, human behavior analysis, and knowledge management. The proposed architecture explicitly addresses critical challenges related to scalability, cognitive bias mitigation, and ethical compliance by introducing algorithmic relevance validation, memory governance, and operational safeguards. These mechanisms establish a robust foundation for future research on continuous learning, sustainability, and multimodal cognitive interaction. **This manuscript constitutes a substantially revised and extended version of the previously released preprint (arXiv:2502.04259, v1), incorporating a formal architectural specification, explicit algorithmic governance mechanisms, clearer integration with large language models, and operational bias mitigation strategies.**

## 1. Introduction to Cognitive AI

### 1.1. Definition

The emulation of human cognitive processing is one of the challenges faced by Artificial Intelligence (AI), this paradigm focuses on an approach that goes beyond the traditional capabilities of AI systems, such as processing textual, graphic, or audible information by answering queries; but also in understanding, learning, and adapting based on previous interactions (Sereati et al., 2020). Therefore, Cognitive AI aims to replicate human cognitive functions, such as memory, learning, and reasoning, to achieve a more natural and effective interaction with users, thereby facilitating the development of a new stage in Human-Computer Interaction (Malyala et al., 2023).

### 1.2. The purpose and scope

The purpose of this Framework is focused on bridging the gap between traditional AI systems, which operate primarily on predefined rules and patterns (Pandey & Bhat, 2023), and a more advanced level of intelligence that simulates human cognitive capabilities; Cognitive AI seeks to provide a structured basis for the standardization of this new type of architecture that allows AI systems not only to access and use previously trained knowledge, but also to accumulate, update and manage the knowledge acquired over multiple interactions, improving personalization and adaptability in their responses (Zheng et al., 2023).

### 1.3. Basic principles of knowledge management and integration

Within this approach, three essential key concepts are highlighted to understand how it works, which will be discussed in depth in the following sections, however each of them will be described in the form of an introduction, we start with "*the context of interaction*" which acts as a long-term memory, which stores and manages persistent information about the user's interactions,  allowing for continuous personalization and a deeper understanding of user preferences (Fuente & Pousada, 2019).

On the other hand, we have the "*key conversational context*" concerning its functions like that of short-term memory, retaining relevant information only during the current session to maintain the coherence and flow of the dialogue (Marri, 2024). Finally, "*the integration of knowledge*" is a fundamental aspect as it allows us to combine the initial training data of the model with live interactions, enabling the AI system to evolve and adapt based on past experiences and new data acquired (Liu et al., 2023).

## 2. Core Components of the Framework

### 2.1. User Interaction Module

The *User Interaction Module* aims to act as the main interface between the AI system and the user, allowing data entry and conversation maintenance, among its main roles is to capture user requests and feedback in real-time, providing a fluid and accessible experience (Meng et al., 2024), therefore, this module will be responsible for initiating and managing interactions, adapting to user behavior to improve communication and foster a friendly and efficient user environment. Among the main features of this module, we propose two combined elements that will provide a robust framework of security, continuity, and coherence of the user experience.

a) *User authentication*, which will guarantee the security and personalization of interactions, the monitoring of the interaction that allows recording and analyzing usage patterns.

b) *Session management*, through which you can facilitate the management of multiple conversations in an orderly manner.

### 2.2. Context of Conversation (*Short-Term Memory*)

Below, the conversation context will be explained in detail, which can be compared to some fundamental characteristics of short-term memory, as it manages the temporal information that is used during interaction sessions, this "*context"* focuses on retaining the details necessary for the current conversation, ensuring that the system can respond in a coherent and contextualized way (Marri, 2024), this type of memory is essential for maintaining fluidity in dialogue and responding appropriately to user questions or comments in a limited time frame (Liu et al., 2023).

A key feature of the "*conversation context"* is its re-establishment at the end of the session, being a design aimed at guaranteeing the efficiency of the system, this because the retained information is only useful during the current interaction and therefore the flow of data within this *context* it includes the capture of immediate responses, real-time updates, and the user's questions, allowing the system to maintain an up-to-date and contextual understanding of the conversation (Finch & Choi, 2024).

### 2.3. Context of Interaction (*Long-Term Memory*)

The other type of memory is the "*interaction context"* which could be represented as the long-term memory of the system, functioning as a repository of persistent information that stores user preferences, historical data and previous interactions, this type of memory is crucial for the personalization of the experience, as it allows the system to retain and use relevant information throughout multiple sessions, this is because by maintaining details about the user's history, the system can offer more accurate answers and adapt its behavior to meet better the user's expectations and needs (Fuente & Pousada, 2019).

The storage of this memory is implemented in a structured database designed to ensure persistence and efficient access, as well as a well-defined synchronization process that allows important data from the "*conversational context"* to be transferred to long-term memory, ensuring that relevant information remains available for future interactions and thus improving the continuity of the user experience (Finch & Choi, 2024).

### 2.4. Knowledge Base Integration

"*Pre-trained model knowledge"* is based on training with large datasets containing general information of the world and complex language patterns, this source of knowledge allows the AI system to answer a variety of questions and perform natural language understanding tasks, providing a solid foundation for interactions,  therefore, the capabilities of the pre-trained model include pattern recognition, logical reasoning, and the generation of contextually appropriate responses (Durt & Fuchs, 2024).

On the other hand, "*dynamic knowledge refreshing"* refers to the ability of the system to evaluate, store, and use new information acquired during user interactions, this process, known as persistent memory updating, ensures that the system remains adaptable and can integrate new data into its knowledge structure, improving the personalization and relevance of its responses (Zheng et al., 2023).

### 2.5. Context Management Database

The "*structure of a unified database"* is a fundamental piece to manage both the conversational context (short-term memory) and the interaction context (long-term memory), this database allows for the efficient storage and retrial of key information, facilitating the updating and synchronization of data as needed, unifying both contexts in a single database ensures that the system can access all relevant information quickly and consistently (Finch & Choi, 2024).

For this reason, "*state persistence mechanisms"* include session persistence, which ensures that data from a session can be maintained and potentially elevated to long-term memory if it is determined to be relevant (Durt & Fuchs, 2024), and context synchronization, which is the process by which data from the current session is evaluated and transferred to long-term storage. They optimize the retention of meaningful information.

### 2.6. Cognitive Processing Modules

Within this module, we have "*logical and analytical processing"* which in the AI system is similar to the functions of the left hemisphere of the human brain, which is responsible for handling structured tasks that require sequential reasoning, data analysis, and language processing (Sholihah, 2022). It is essential for performing logical operations and providing precise and detailed answers that require a high level of technical and structured understanding.

On the other hand, "*creative and pattern recognition processing"* is comparable to the functions of the right hemisphere, which manages tasks that require a broader understanding of the context, allowing for the generation of creative responses and the interpretation of complex data (Aberg et al., 2016). The system uses this type of processing to adapt to different contexts and make inferences, offering answers that go beyond the literal and encompass richer and more nuanced interpretations.

### 2.7 Architecture Specification: Interfaces, Data Objects, and Control Plane

The Cognitive AI Framework proposed is organized as a layered and interface-driven cognitive architecture that explicitly separates information processing from governance and control mechanisms. This design follows a dual-plane principle: a data plane, responsible for perception, retrieval, reasoning, and generation. And a control plane, responsible for memory governance, privacy enforcement, ethical constraints, and auditability, in line with contemporary discussions on responsible and explainable AI (Floridi et al., 2018).

The core data objects manipulated by the system include: (a) UserMessage, representing the incoming interaction with metadata; (b) SessionState (Short-Term Memory, STM), which maintains the active conversational context, extracted facts, goals, and risk flags; (c) MemoryItem (Long-Term Memory, LTM), which stores persistent knowledge units enriched with vector embeddings, domain tags, provenance, confidence scores, and retention time-to-live (TTL); and (d) PolicyProfile, which encodes privacy rules, consent flags, and safety thresholds.

From an architectural perspective, this approach differs from classical symbolic cognitive architectures such as SOAR (Newell, 1990) or ACT-R (Anderson, 2004) by adopting a hybrid representation in which symbolic descriptors and distributed vector representations coexist. The data plane is composed of four main functional modules: extraction, retrieval, reasoning/generation, and post-processing. The control plane supervises all persistence and access decisions through a persistence gate, which evaluates candidate memory items against relevance, sensitivity, and policy constraints. This explicit governance layer is critical in long-term human–AI interaction scenarios, where uncontrolled accumulation of user data may lead to ethical, legal, and epistemic risks (Floridi et al., 2018).

### 2.8 Integration with Large Language Models (LLMs): Retrieval, Memory, and Tooling

This framework is designed to integrate large language models (LLMs) as a frozen generative core, without modifying their internal parameters through user interaction. Instead, it adopts a retrieval-augmented generation (RAG) strategy (Lewis et al., 2020), combined with a governed external memory system.

Three layers can be distinguished in this process. First, the LLM core is responsible for linguistic generation and reasoning, leveraging transformer-based architectures (Vaswani et al., 2017). Second, the retrieval layer queries the long-term memory (LTM) and approved knowledge sources to assemble a contextually relevant evidence set. Third, the memory governance layer decides what information may be persisted, updated, or discarded based on relevance validation, privacy policies, and ethical constraints.

Operationally, user input updates the STM, candidate memory elements are extracted, relevant evidence is retrieved from LTM, and a structured prompt is assembled. The LLM then generates a response, which is subsequently subjected to post-generation safety and consistency checks. This design allows the system to benefit from the expressive power of LLMs while avoiding the well-known risks associated with ungoverned long-term data accumulation and opaque adaptation processes (Floridi et al., 2018).

## 3. Data Flow and Process Logic

### 3.1. Interaction Flow

The interaction flow describes the structured process by which information is handled from the moment the user initiates a query until the update of the system memories, it is important to emphasize that this flow begins with the user's search input, where the request is initially processed in the context of conversation (*short-term memory*),  being here where the system analyzes and responds based on the temporal data of the current session (Finch & Choi, 2024), therefore, this management of the conversation context is essential to maintain the coherence and relevance of the responses during the Human-AI interaction.

It is also important to consider that within this flow process, as the session progresses, an evaluation of the relevance of the interaction in the long term is carried out; During this process, the system determines if the information captured during the session should persist beyond the current interaction, therefore, if under this premise the information is identified as relevant, it is transferred to the context of interaction (*long-term memory*), thus ensuring that the system can use this data in future interactions to improve personalization and user understanding.  Below, we will explain these two concepts in detail (Fuente & Pousada, 2019; Liu et al., 2023).

#### 3.1.2. Relevance Validation

As we explained earlier, once the system identifies information potentially relevant, the system proceeds to validate it using methods based on algorithms and predefined rules. For this, it applies a data weighting system, assigning a score to each element of information according to its relative importance (Zheng et al., 2023).

If this score exceeds a set threshold, the information is classified as suitable to be stored in long-term memory. Another critical step is the evaluation of temporal persistence, which is to determine if the data has a prolonged impact or is simply ephemeral. For example, a one-time reminder of a meeting may not meet persistence criteria, while details related to recurring events or personal preferences are often validated for long-term storage (Finch & Choi, 2024).

To better understand this process, it is important to understand that the validated information is compared with the existing data in the context of interaction, so if it complements, updates, or improves prior knowledge without redundancy, its relevance is confirmed, guaranteeing efficient memory management (Meng et al., 2024).

### 3.1.3. Decision Flow for Persistence

In the case of persistence decisions, the system follows a structured flow to ensure that only significant information is retained in long-term memory. This is because, during each session, the captured information is initially stored in short-term memory, where it is subjected to an exhaustive analysis using the criteria (Torrentira, 2024).

In this analysis, the system determines whether the data meets the requirements to be retained, if Information is considered relevant and persistent, transferred to the context of interaction, and integrated as part of long-term memory (Christoph & Fuchs, 2024). Conversely, if the data fails to meet the established criteria, it is deleted after the session to prevent system overload and preserve operational efficiency. This approach ensures that the system not only manages memory effectively but also enhances the user experience by retaining only information that offers meaningful long-term value.

### 3.1.4 Algorithmic Relevance Validation and Memory Persistence

To operationalize selective memory formation, the framework introduces an explicit algorithmic gate between short-term and long-term memory. This approach is conceptually aligned with cognitive theories that distinguish working memory from long-term memory consolidation (Anderson, 2004), while also addressing ethical and epistemic concerns in artificial systems (Floridi et al., 2018)

**Algorithm 1:**

Relevance Validation and Memory Persistence (STM → LTM)

```
 Inputs:
  m_t: user message at turn t
  S_t: session state (STM) at turn t
  M: long-term memory store (LTM)
  P: policy profile (privacy and ethics constraints)

Outputs:
  r_t: system response
  M': updated LTM
1) S_t ← UpdateSTM(S_{t-1}, m_t)
2) candidates ← ExtractCandidates(S_t)
   # facts, stable preferences, recurring entities, constraints, corrections
3) evidence ← RetrieveEvidence(m_t, S_t, M)
4) draft ← GenerateResponse(m_t, S_t, evidence)
5) scored ← []
```

```
   For each c in candidates:
       score = w1*Stability(c) + w2*Utility(c) + w3*Recurrence(c, S_t)
               + w4*Novelty(c, M) - w5*Sensitivity(c, P) - w6*Conflict(c, M)
       If ViolatesPolicy(c, P): continue
       scored.append((c, score))

6) writes ← { c | (c,score) in scored and score ≥ θ_persist }
7) M' ← UpsertMemory(M, writes)
   # merge duplicates, attach provenance, set TTL, store embeddings and tags
8) r_t ← PostProcess(draft)

Return r_t, M'
```

This mechanism ensures that only information that is stable, useful, non-sensitive, and non-contradictory is promoted to long-term memory. In contrast to purely generative systems, which may implicitly internalize user data in opaque ways, the proposed framework makes memory formation an explicit, auditable, and reversible process.

### 3.2. Use of Knowledge

In a Cognitive AI system, knowledge is used through the combination of its accessibility and pre-training, through its dynamic integration based on previous interactions, with pre-trained knowledge being the fixed base of information that allows the AI system to have a solid starting point, which is key to providing immediate and reliable answers (Marri, 2024).

It is also important to emphasize that pre-trained knowledge has a limitation, which is that it cannot be updated automatically, therefore, it is important to consider that this knowledge is derived from large volumes of data with which the model was initially trained, allowing it to understand natural language and effectively answer a variety of general questions (Durt & Fuchs, 2024).

In addition to pre-trained knowledge we have dynamic knowledge which is the information that is continuously adapted and updated with new interactions, which allows the AI system to personalize and improve its responses over time, this type of knowledge is consulted and updated in real-time as the system interacts with the user,  ensuring that responses are increasingly relevant and personalized (Zheng et al., 2023), with this updating process being a reinforcement in the system's ability to learn from previous experiences and apply that learning in future interactions, optimizing the personalization and accuracy of responses.

To conclude this section, we must mention that both types of knowledge interact synergistically to optimize the responsiveness of AI systems, combining the robustness of a pre-trained knowledge base, which offers a comprehensive and grounded understanding of language and pre-established patterns, with the adaptability of dynamic knowledge, which allows continuous updating and contextual personalization (Meng et al., 2024).

**3.3. Decision Framework**

Within the data flow, the decision framework is a critical structure that guides the process of identifying, evaluating, and storing the data generated during the interaction with the user. This framework determines whether the information captured during the current session is significant enough to be retained in long-term memory (Finch & Choi, 2024).

This process begins with the identification of relevant data, followed by an evaluation in which it is decided whether that data meets the persistence criteria. For this reason, if the decision is affirmative, the information is transferred to the context of interaction, ensuring that it is available for future consultation and use (Fuente & Pousada, 2019).

To represent this process clearly and visually, we have the following flowchart which describes the complete cycle from user input, management in the conversation context, persistence assessment, to long-term memory update, this decision framework also provides a transparent approach to how data is handled,  ensuring that the transition between short-term and long-term memory is carried out efficiently and according to predefined criteria that improve the user experience and the functionality of the system (Christoph & Fuchs, 2024; Torrentina, 2024).

**Figure 1**

Cognitive AI Data Flow. *Own Creation*

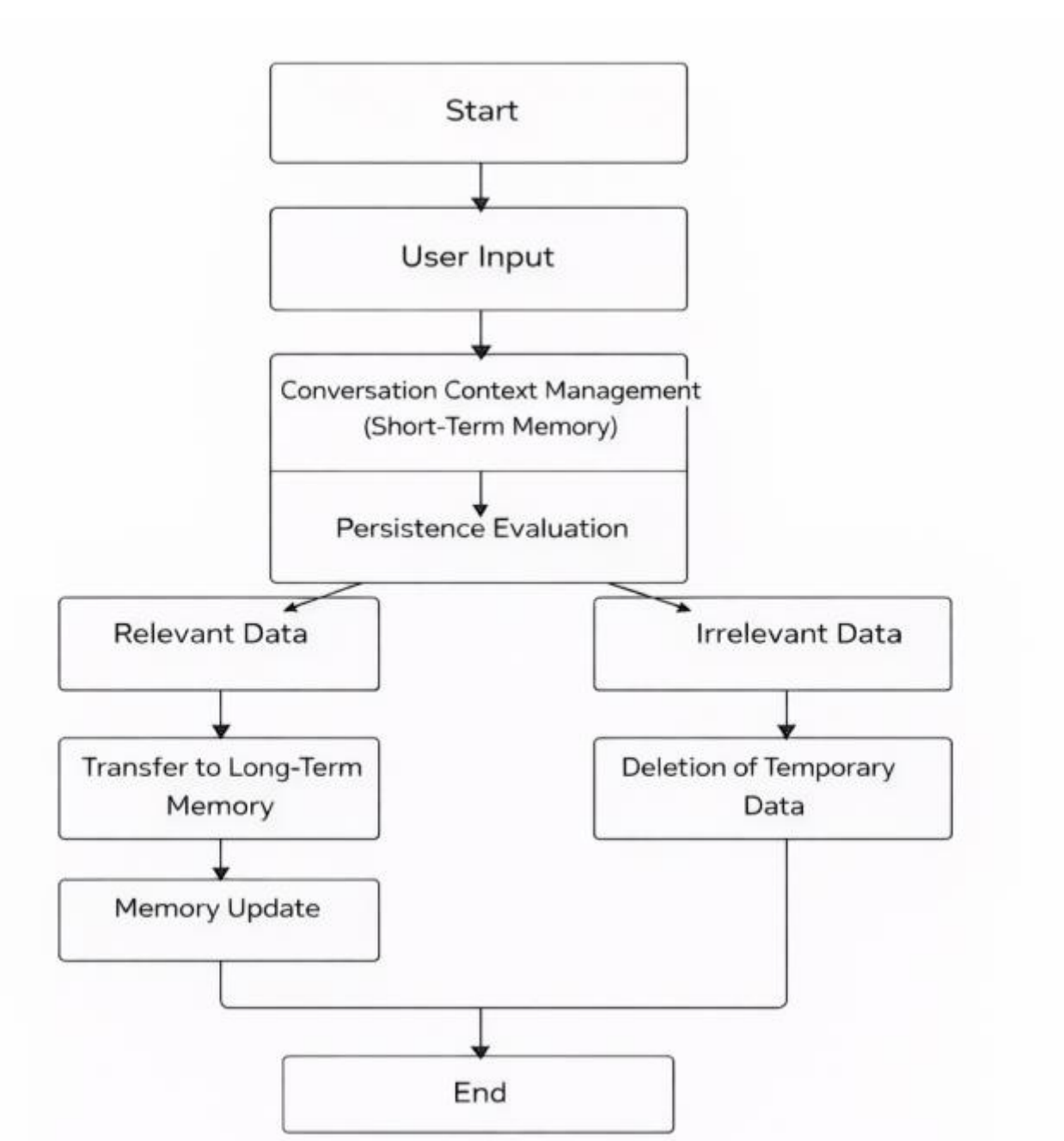

## 4. Cognitive Comparative Analysis

### 4.1. Human Cognitive Processes vs. AI Model Capabilities

This analysis seeks to understand how human cognitive processes and the capabilities of an AI model provide a deep understanding of the similarities and differences in an information management environment, as human beings and their cognitive processes, such as perception, reasoning, memory, and decision-making, take place in the brain.  which can learn from experience, adapt to new situations, and apply accumulated knowledge to solve problems effectively (Durt & Fuchs, 2024).

For this reason, this Framework aims to seek a deeper understanding of how Cognitive AI technology can replicate biological functions through algorithms and models that mimic the structure of a neural network, capturing complex patterns of data to respond intelligently and adaptively (Zheng et al., 2023), below, a matrix that explains this comparative analysis.

**Table 1**

Benchmarking Matrix: Traditional AI and Cognitive AI. *Own Creation*

| Process | Traditional AI | Cognitive AI |
|---|---|---|
| *User Input* | The request is received, but it does not retain context. | The request is received and stored in short-term memory. |
| *Conversation Management* | Responds according to pre-programmed patterns. | Personalize the response based on previous interactions in the same session. |
| *Data Storage* | Information is deleted at the end of the session. | The information is evaluated and, if relevant, |
| | | stored in long-term memory. |
| *Adaptability* | No personalization in future sessions. | AI enhances and adapts its responses in future interactions. |

Unlike what is explained with the human brain, in an AI model, the pre-trained model acts as the brain nucleus, containing a vast repository of general knowledge and language patterns acquired during its initial training with large amounts of data (Torrentira, 2024), this model allows the AI to perform tasks such as language understanding,  text generation and response to questions consistently and quickly (Chen et al., 2024).

However, although this nucleus mimics some capabilities of the human brain, such as the Logical reasoning and comprehension (Ye et al., 2023), are limited in their ability to improvise outside of the patterns and data on which they were trained, unlike the natural flexibility and adaptability of the human mind (Finch & Choi, 2024).

### 4.2. Analogies of Short-Term and Long-Term Memory

A useful comparison between AI models and the human brain is the functioning of memory, because, in humans, short-term memory, or working memory, temporarily retains information to process immediate tasks, such as remembering an address or following a conversation, this storage capacity being crucial to maintain consistency in interactions and handle multiple pieces of information at the same time (Marri,  2024), below, a matrix that explains this comparative analysis.

**Table 2**

Benchmarking Matrix: Human Memory and AI Short-Term Memory. own creation

| Aspect | Human Memory (Short Term) | AI Memory (Short Term) |
|---|---|---|
| *Information retention* | Temporarily retains information | Stores information relevant to the current session |
| *Memory Duration* | Second to minutes | During the session, it is deleted at the end of the session. |
| *Storage capacity* | Limits around seven elements | Depends on the capacity of the system |
| *Flexibility and Upgrade* | Flexible but susceptible to interference | Updatable in real-time, but no post-session retention |
| *Using Interactions* | Process immediate tasks such as remembering an address | Maintain consistency in the conversation |
| *Multitasking* | Can handle multiple items simultaneously with effort | Processes multiple requests based on model capacity |

However, on the other hand, long-term memory will allow human beings to store information in a more lasting way while being involved in learning and accumulating knowledge over time, which allows people to apply past experiences to solve future problems (Fuente & Pousada, 2019).

In an AI system, these concepts are reflected in the methods called "*Contexts*", being the context of conversation, the analogy of short-term memory, and the context of interaction of long-term memory. To better understand these concepts, they will be explained below. In the case of the conversational context, it retains data for the duration of a session, like how working memory handles information temporarily. Once the session ends, this memory is restarted unless relevant information is identified that must be stored in long-term memory (Liu et al., 2023).

When stored in long-term memory, the interaction context comes into play, which functions like long-term memory, storing accumulated data that can be consulted and used in future interactions, which allows AI to improve its ability to personalize and retain important information persistently (Meng et al., 2024).

Finally, we must emphasize that, although this analogy is of utmost relevance in Cognitive AI as it will help us understand how AI models are designed to mimic the structure of human cognition, there are some limitations in terms of depth and flexibility in adaptive learning that should be explored in the future (Christoph & Fuchs, 2024).

### 4.3 Positioning Against Cognitive Architectures and Neuro-Symbolic Systems

Classical cognitive architectures such as SOAR (Newell, 1990) and ACT-R (Anderson, 2004) provide formally grounded models of cognition based primarily on symbolic or hybrid symbolic–subsymbolic representations. These architectures have been highly influential in modeling human problem-solving and procedural reasoning but are not designed for open-ended, language-intensive, and continuously evolving interaction scenarios.

In contrast, modern LLM-based systems, built on transformer architectures (Vaswani et al., 2017), exhibit strong linguistic and inferential capabilities but typically lack explicit, governed long-term memory and transparent mechanisms for knowledge persistence (Lewis et al., 2020). The proposed Cognitive AI Framework positions itself as a bridging architecture between these two paradigms: it does not attempt to replace formal cognitive theories, nor to redefine the internal mechanisms of LLMs, but instead introduces a governed interaction and memory layer that enables cognitively coherent long-term human–AI interaction.

The main contribution lies in the explicit coupling of (a) short-term and long-term memory, (b) algorithmic relevance validation, and (c) ethical governance mechanisms, thereby transforming interaction history into a managed cognitive resource rather than an uncontrolled data residue (Floridi et al., 2018).

## 5. Technical Implementation and Infrastructure

### 5.1 Tech Stack

The technological stack of a Cognitive AI framework is made up of various tools and technologies designed to handle and process large volumes of data efficiently. Among them are structured databases, which are essential to store and manage both the interaction context and the conversation context, ensuring the persistence of data and its rapid recovery (Marri, 2024).

These databases include certain technologies such as PostgreSQL, NoSQL, and MongoDB, which offer flexibility in managing unstructured and structured data (Aguilar Vera et al., 2023), It is also important to emphasize the role of cloud infrastructure, such as that provided by AWS, Microsoft Azure, or Google Cloud, which facilitates the scalable storage and processing capacity necessary to handle complex AI operations, allowing fast and secure access to computing resources (Torrentira, 2024).

On the other hand, we have data management which is supported by container orchestration technologies, such as Kubernetes, which ensure efficiency in the management of applications and their deployment, in addition, to these technologies we have real-time data processing tools and batches, such as Apache Kafka and Spark, which guarantee the updating and synchronization of the long-term and short-term memory of AI (Aruna, & Gurunathan, 2024).

As we have been able to observe in the discussions, this technological stack allows us to understand in an easier and summarized way how a robust and scalable environment works, which can allow a Cognitive AI system to adapt to the workload and requirements in human-AI interaction (Meg et al., 2024).

### 5.2. Scalability and Performance Considerations

Scalability plays a key role in the implementation of a Cognitive AI framework, as it allows the system to handle an increase in the number of users or the complexity of queries without losing performance, to achieve this, distributed architectures are implemented that use the capacity for horizontal scaling, i.e., the addition of more nodes to the system to distribute the workload (Zheng et al., 2023).

This information management strategy is complemented by database optimization through partitioning techniques and the use of caching to accelerate data retrieval through the use of monitoring tools, such as Prometheus and Grafana, which are used to monitor system performance and detect bottlenecks in real-time (Chieu & Zeng, 2008).

Performance is addressed by optimizing processing algorithms and implementing AI models that minimize latency in the generation of responses, which must be able to perform fast inferences, even when accessing large volumes of long-term memory data, another complement is parallelization techniques, and the implementation of GPUs or TPUs (processors designed for machine learning operations) which help maintaining efficiency in processing large amounts of data and simultaneous tasks, ensuring a smooth, and seamless user experience (Pandey & Bhat, 2023).

### 5.3. Data Security and Privacy Protocols

As for data security and privacy, this is an issue of relevance in a system of Cognitive AI. Due to the handling of personal and sensitive data, the protocols that must be used guarantee the protection of information in all stages of data storage and transfer (Torrentira, 2024). In addition, it is important to emphasize that encryption technologies, such as TLS (Transport Layer Security), ensure the integrity of data during transmission, while techniques such as data-at-rest techniques, such as AES (Advanced Encryption Standard), protect the information stored in databases (Zhang et al., 2023; Hazra et al., 2024), for this reason, the authentication and authorization of users through access control systems, such as OAuth and two-factor authentication (2FA) protocols, strengthen the security of interactions and access to the system (Vilasini et al., 2024).

Data privacy is ensured by implementing policies in compliance with data protection regulations, such as the GDPR (General Data Protection Regulation) and CCPA (California Consumer Privacy Act), with these policies being clear rules on how personal data should be collected, processed, and stored (Bakare et al., 2024).

Furthermore, data anonymization and pseudonymization techniques are also integrated to protect the user's identity in long-term memory storage and during persistent memory updates. These approaches ensure that AI can operate safely and ethically, protecting user information while optimizing the personalization and relevance of responses (Christoph & Fuchs, 2024).

## 6. Applications and Use Cases

### 6.1. Practical Implementations

The practical applications of a Cognitive AI framework are wide, highlighting its impact on sectors such as education and human behavior. In the educational field, Cognitive AI can make it possible to create adaptive learning platforms that respond to the individual needs of students. These platforms could adjust the content and level of difficulty based on progress and learning style, acting as personalized tutors that provide real-time feedback and specific recommendations, thus improving the effectiveness of the educational process (Marri, 2024).

On the other hand, in cyberpsychology, Cognitive AI can facilitate personalization in digital psychological and therapeutic interventions, since these systems can adapt to the emotions, behaviors, and needs of the user in real-time, offering emotional support, early detection of psychological problems, and personalized guides for well-being (Liu et al., 2023).

Therefore, these two application examples demonstrate how Cognitive AI has the potential to transform key sectors through personalization and adaptability, significantly improving the user experience, as is the case with personalized assistants based on Cognitive AI, improving traditional assistants, through a deeper understanding in the management of user preferences over time to offer a more natural interaction and relevant (Meng et al., 2024).

Therefore, today and in a holistic way, Cognitive AI can also be a tool applicable to business decision-making, by processing large volumes of data, identifying patterns and generating recommendations to optimize business strategies, improving customer service and automating complex processes, to obtain efficiency, by reducing costs and enabling organizations to adapt quickly to changes in the work environment.

### 6.2. Advantages and Limitations

The strengths in this research work were able to identify some advantages of Cognitive AI especially in its ability to offer a high level of personalization and adaptability, due to the integration of long-term and short-term memory, since these systems can learn from past interactions and apply that knowledge to improve future responses and decisions (Zheng et al., 2023).

Bringing this customization to a more satisfactory user experience, since the system adapts and responds more precisely to individual needs, with the capabilities that AI has to process and analyze large volumes of data in real-time time allowing faster and more informed decision-making.

However, some inherent limitations and challenges have been identified in this study that must be considered, such as the handling of large amounts of data, which requires advanced infrastructure to repel scalability and efficiency problems, in addition to one of the most important challenges that is the risk of cognitive biases, which can arise if the training data or previous interactions contain implicit biases which could influence the decision-making of AI and affect the fairness and objectivity of its responses (Fuente & Pousada, 2019).

Therefore, to mitigate these risks, it is critical to implement training and evaluation strategies that promote transparency and reduce bias, as well as ensure that AI operates ethically and complies with privacy and data protection regulations, such as GDPR and CCPA (Bakare et al., 2024).

### 6.3 Bias Mitigation and Ethical Safeguards: Operational Controls

Concerns regarding bias, opacity, and misuse of AI systems have been widely documented in the literature (Bender et al., 2021; Floridi et al., 2018). Rather than treating these issues as purely normative, the proposed framework embeds concrete operational safeguards directly into its architecture.

At the data and memory level, each memory item is stored with provenance metadata, confidence scores, and expiration parameters, reducing the risk of long-term entrenchment of erroneous or context-dependent information. At the retrieval and generation level, the system supports diversity-aware retrieval strategies and uncertainty calibration, mitigating the tendency of generative models to produce overconfident or biased outputs (Bender et al., 2021).

At the governance level, user consent profiles, right-to-forget mechanisms, and audit logs ensure that memory persistence remains legally and ethically defensible. In this sense, ethical principles are not merely stated but are translated into enforceable technical constraints, consistent with current frameworks for responsible AI governance (Floridi et al., 2018).

## 7. Conclusion and Future Prospects

### 7.1. Summary of the potential of Cognitive AI

The Cognitive AI framework represents a significant advance in the integration of technologies that mimic and optimize human cognitive abilities. by combining short- and long-term memory with a structured storage infrastructure and advanced cognitive processing, this framework allows a more natural and personalized interaction between the user and the AI.

It is also important to emphasize that the robustness of the system in terms of personalization and contextual learning positions it as an essential tool in the new era of artificial intelligence, its potential lies in its ability to adapt and evolve with each interaction, offering more accurate and relevant responses over time, therefore, this adaptability improves the user experience in applications ranging from cyberpsychology, educational environments and different business solutions (Zheng et al., 2023).

### 7.2. Next steps in research and development

Although the current capabilities of the Cognitive AI framework are extensive, the next steps in research and development should focus on further improving its learning capacity and adaptability, a key direction identified in this study is the future work of continuous learning algorithms that allow Cognitive AI to continue learning and adjusting without the need to be trained from scratch. which would increase their ability to respond to new and changing situations. We also identified the need to implement mechanisms to minimize cognitive biases and improve equity in decision-making, ensuring that the system maintains a high ethical and fair standard in its responses and actions.

Another relevant issue is the use of lighter and more optimized models that can contribute to the generation and implementation of more systems based on Cognitive AI with applications to mobile phones and IoT devices (Rashid et al., 2022), in addition to the integration of advanced perception and multimodality capabilities (e.g., text blending,  voice and image), which would expand the applicability of the framework to new scenarios and use cases, ensuring that the growing expectations of AI users are met.